\documentclass[twocolumn,conference,letterpaper]{IEEEtran}
\usepackage[left=0.75in,right=0.75in,top=1in,bottom=0.75in]{geometry}

\usepackage{cite}
\usepackage{amssymb,amsmath,latexsym,amsfonts,amsthm,mathtools}

\usepackage{graphicx}
\usepackage{float,subcaption}
\usepackage{textcomp}
\usepackage{xcolor}
\definecolor{darkpastelpurple}{rgb}{0.59, 0.44, 0.84}
\usepackage{hyperref}

\usepackage{algorithm}
\usepackage{algpseudocode}

\hypersetup{colorlinks, breaklinks, citecolor=blue, linkcolor=blue, urlcolor=blue}
\usepackage[nameinlink]{cleveref}
\Crefformat{figure}{#2Fig.~#1#3}
\Crefmultiformat{figure}{Figs.~#2#1#3}{ and~#2#1#3}{, #2#1#3}{ and~#2#1#3}

\usepackage{tikz}
\usetikzlibrary{automata, shapes, arrows, calc, arrows.meta, fit, positioning}

\theoremstyle{plain}
\newtheorem{theorem}{Theorem}
\newtheorem{lemma}{Lemma}

\newtheorem*{problem*}{Problem}
\theoremstyle{remark}

\theoremstyle{definition}

\newcommand{\sign}{\mathrm{sign}}
\usepackage{mathrsfs}

\usepackage{newtxmath}
\usepackage{booktabs}
\usepackage{duckuments}

\IEEEoverridecommandlockouts

\begin{document}
	\title{3D Guidance Law for Flexible Target Enclosing with Inherent Safety}
	\author{Praveen Kumar Ranjan, Abhinav Sinha,~\IEEEmembership{Senior Member,~IEEE}, Yongcan Cao,~\IEEEmembership{Senior Member,~IEEE}
\thanks{P. K. Ranjan and Y. Cao are with the Unmanned Systems Lab, Department of Electrical and Computer Engineering, The University of Texas at San Antonio, TX- 78249 USA. Emails: praveen.ranjan@my.utsa.edu, yongcan.cao@utsa.edu.\newline A. Sinha is with the GALACxIS Lab, Department of Aerospace Engineering and Engineering Mechanics, University of Cincinnati, OH- 45221 USA. Email: abhinav.sinha@uc.edu.}
}

	\maketitle
	\thispagestyle{empty}
	
	\begin{abstract}
In this paper, we address the problem of enclosing an arbitrarily moving target in three dimensions by a single pursuer while ensuring the pursuer’s safety by preventing collisions with the target. The proposed guidance strategy steers the pursuer to a safe region of space surrounding and excluding the target, allowing it to maintain a certain distance from the latter while offering greater flexibility in positioning and converging to any orbit within this safe zone. We leverage the concept of the Lyapunov Barrier Function as a powerful tool to constrain the distance between the pursuer and the target within asymmetric bounds, thereby ensuring the pursuer’s safety within the predefined region. Further, we demonstrate the effectiveness of the proposed guidance law in managing arbitrarily maneuvering targets and other uncertainties (such as vehicle/autopilot dynamics and external disturbances) by enabling the pursuer to consistently achieve stable global enclosing behaviors by switching between stable enclosing trajectories within the safe region whenever necessary, even in response to aggressive target maneuvers. To attest to the merits of our work, we conduct experimental tests with various plant models, including a high-fidelity quadrotor model within Software-in-the-loop (SITL) simulations, encompassing various challenging target maneuver scenarios and requiring only relative information for successful execution.
	\end{abstract}
	
	\begin{IEEEkeywords}
    Multiagent Systems, Target Enclosing Guidance, Motion Planning, Safety and Path Planning.
    \end{IEEEkeywords}

\section{Introduction}    
In recent times, autonomous vehicles have gained significant prominence in a range of tasks, including reconnaissance, surveillance, crop and forest monitoring, and search and rescue missions \cite{9121685, 9697374,10634571, 8392366, 8613865,doi:10.2514/1.G007057}. Most of these tasks share a common theme of behavior where an autonomous vehicle known as the pursuer monitors an object of interest known as the target. The pursuer achieves this by maintaining a specific distance or proximity to the target, which can be stationary or dynamic. The pursuer's ability to achieve a stable motion around the target, thereby maintaining fixed desired proximity, is commonly referred to as target enclosing, target circumnavigation, or target encirclement \cite{10251969}. One of the earliest approaches to address the target enclosing problem involved utilizing cyclic pursuit by agents to achieve a circular formation around the target \cite{KIM20071426, MARSHALL20063}. In another approach known as vector field guidance, the pursuer enclosed the target utilizing vector fields designed to create a limit cycle or a periodic orbit of the desired shape around the target \cite{doi:10.2514/1.30507, GAO2022107800,doi:10.2514/1.G002281,targettrack,9523776}. Several other methods have been developed for trapping one or more targets in circular orbits based on different guidance strategies while also focusing on reducing the communication effort \cite{doi:10.2514/1.G001421,CAO2015150,doi:10.2514/1.G002707,6705614,LAN2010381,doi:10.2514/1.G006403}. For example, only range measurements were utilized in \cite{CAO2015150}, while only bearing information was utilized in \cite{doi:10.2514/1.G001421,doi:10.2514/1.G002707}. In \cite{LAN2010381,doi:10.2514/1.G006403}, only relative information (range and bearing measurements) was utilized for the development of target encirclement laws, whereas the work in \cite{10251969,doi:10.2514/1.G006957,9924233,doi:10.2514/1.G007539,ranjan2024selforganizing,doi:10.2514/6.2024-0124} addressed a general target enclosing problem using relative information.

The above-mentioned works developed target-enclosing laws with restrictions on the target's motion (stationary \cite{GAO2022107800,targettrack,doi:10.2514/1.G001421,CAO2015150,9924233,doi:10.2514/1.G007539}, moving target \cite{doi:10.2514/1.30507,doi:10.2514/1.G002281,doi:10.2514/1.G006403,10251969} or maneuvering target \cite{9523776,doi:10.2514/6.2024-0124}). Such simplifications make it possible to obtain a tractable feedback controller but may require information on the target maneuver, necessitating
communication between the target and the pursuer. For example, in \cite{rendev2013}, the authors assumed the target to be moving in a straight line, while the authors in \cite{doi:10.2514/1.G006403} developed separate laws when dealing with constant acceleration, constant velocity, or stationary target. Further, despite the practical importance of 3D engagements, there has been limited focus on the development of laws for target enclosing in 3D space.  In \cite{10015886}, authors utilized a behavioral model to develop a 3D guidance algorithm to enclose a target in a GPS-denied environment based on camera images. The authors in \cite{10092932} developed target tracking laws for a pursuer with nonholonomic kinematics, orbiting on a path depending on the target's heading and motion, and provided experimental results for an actual airship. Recently, the works in \cite{9924233,doi:10.2514/1.G007539} have developed 3D guidance laws to enclose targets within arbitrary shapes, providing greater flexibility to the pursuer to move freely. It is important to mention that most of the works mentioned above decouple the 3D engagement into two planar engagements or assume simplified target maneuvers, which restricts the movement of the pursuer and may diminish the performance of the guidance law by limiting the allowable paths taken by the pursuer. Additionally, these works primarily focus on the pursuer reaching a desired fixed proximity around the target. However, in unknown environments, the pursuer may not be able to precisely track its position at the desired proximity due to unmodeled dynamics, disturbances, or aggressive target maneuvers, potentially leading to a collision between the pursuer and the target or a loss of stable target enclosing behavior. Therefore, a robust guidance strategy must allow the pursuer certain flexibility to converge within a predefined region around the target rather than aiming for a singular orbit to enclose it.

This letter is motivated by the need to develop a generalized 3D target enclosing law for the pursuer that provides safety guarantees and robustness against arbitrary target maneuvers based solely on relative information. The contributions of this letter include developing guidance laws that enable the pursuer to enclose a mobile target while providing the flexibility to converge to a broader spectrum of relative positions, yielding a multitude of stable enclosing paths. We refer to this capability of the pursuer as \textbf{flexible target enclosing}, as it offers multiple stable trajectory options or a larger set of relative positions while enclosing the target. By ensuring that the target's position and the region to which the pursuer converges never intersect, we can guarantee the pursuer's safety. Additionally, our approach provides inherent robustness against target maneuvers (or other sensitivities, such as wind disturbances and modeling/measurement errors) by allowing the pursuer to smoothly transition between stable enclosing orbits whenever necessary. This makes it suitable for a wider range of operating conditions, such as agile target maneuvers or different classes of UAVs. Finally, we demonstrate the practicality of our guidance law by implementing it on a high-fidelity UAV model via Software-in-the-Loop simulations, providing stronger confidence in the robustness of the guidance law.
\section{Problem Formulation}
In this section, we present a brief overview of the pursuer-target kinematic model and discuss the problem addressed in this paper.
\subsection{Pursuer-Target relative kinematics}
\begin{figure*}[ht!]
	\begin{subfigure}[t]{0.59\linewidth}
		\centering
		\includegraphics[width=\linewidth]{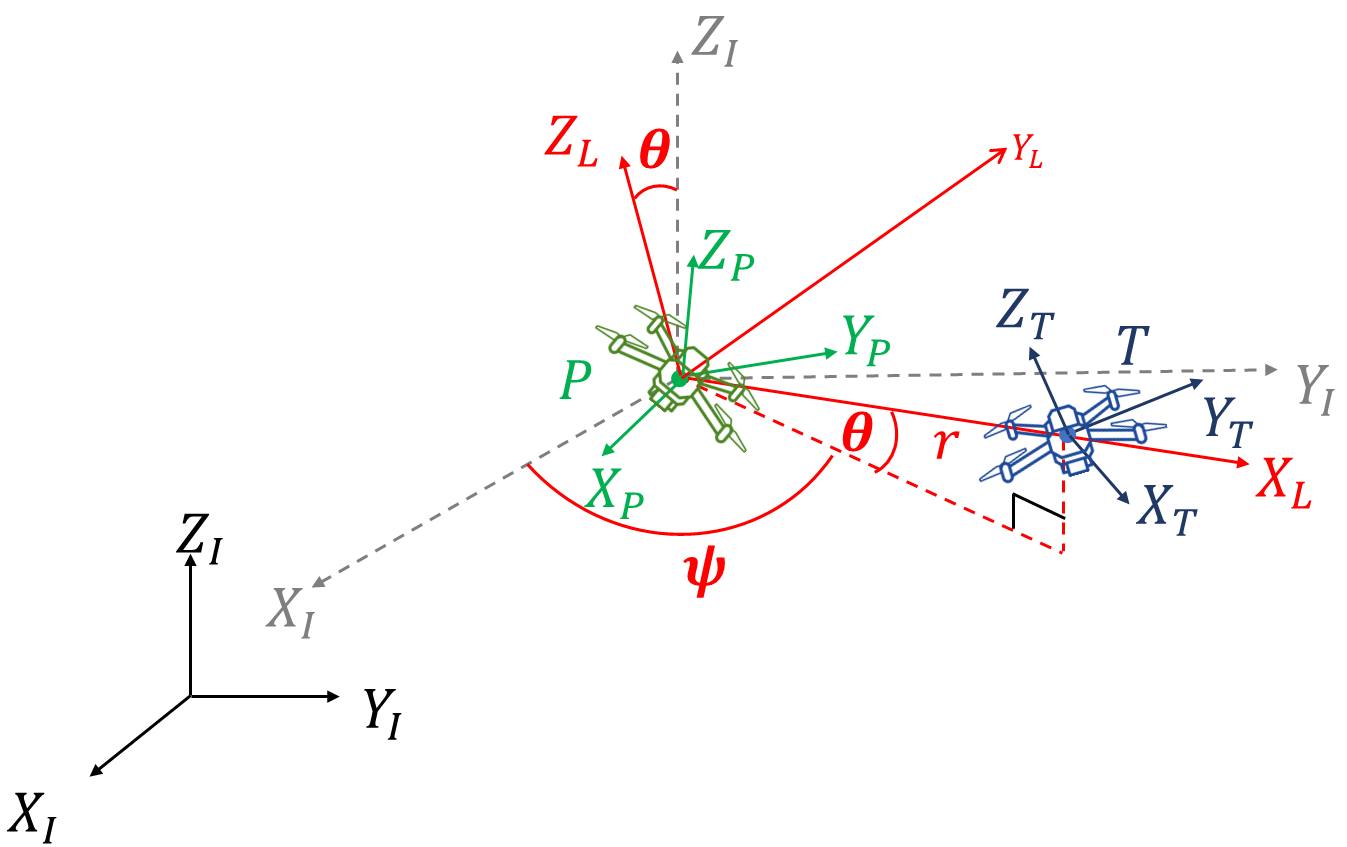}
		\caption{Inertial frame of reference.}
		\label{fig:iner_enggeom}
	\end{subfigure}
	\begin{subfigure}[t]{0.40\linewidth}
		\centering
		\includegraphics[width=\linewidth]{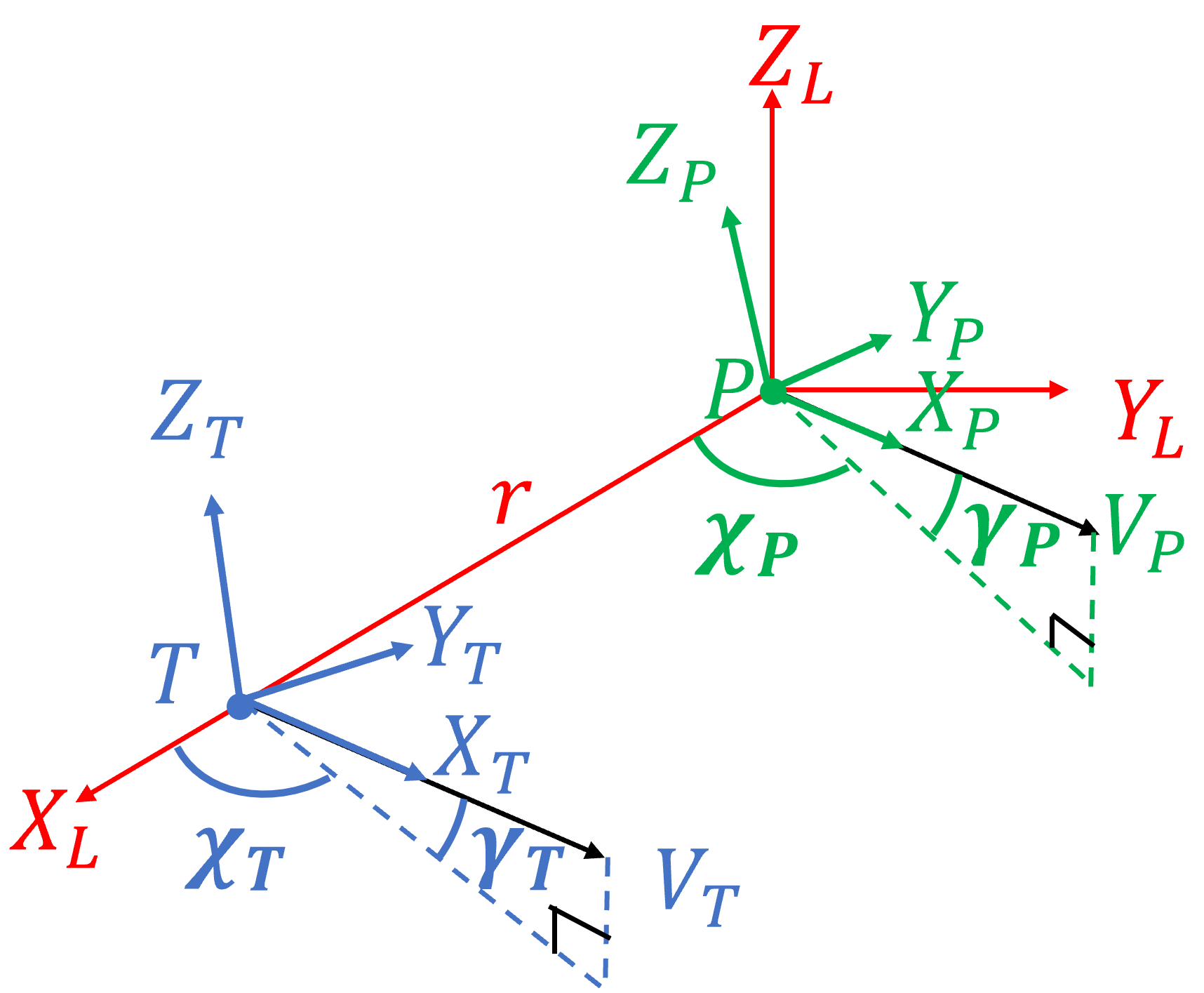}
		\caption{LOS frame of reference.}
		\label{fig:enggeom}
	\end{subfigure}	
 	\caption{Engagement geometry.}
	\label{fig:eeee}
\end{figure*}

Consider the three-dimensional engagement between a pursuer (P) and a target (T) in the inertial frame of reference $F_I$  (denoted by mutually orthogonal $\{X_I, Y_I, Z_I\}$ axes), as illustrated in the \Cref{fig:iner_enggeom}. The pursuer and the target can move at different speeds, denoted by $V_P$ and $V_T$, respectively, while they are separated by a distance $r$. By rotating $F_I$ through azimuth line-of-sight (LOS) angle $\psi \in (-\pi, \pi]$ and elevation LOS angle $\theta \in [-\pi/2, \pi/2]$, we obtain the pursuer-target LOS frame of reference $F_L$, denoted by mutually orthogonal $\{X_L, Y_L, Z_L\}$ axes. $F_L$ has the $X_L$ axis aligned along the line-of-sight (LOS) from the pursuer to the target. Further, the body-fixed frame of reference for the pursuer and target are denoted by $F_P$ and $F_T$, respectively (denoted by mutually orthogonal $\{X_i, Y_i, Z_i\}$ axes where $i \in \{P,T\}$ represents the pursuer and the target). The engagement between the vehicles in $F_L$ is depicted in \Cref{fig:enggeom}, where $V_P$ and $V_T$ are aligned along the $X_i$ axis of their respective body-fixed frame. $F_P$ is determined by rotation of $F_L$ through pursuer azimuth angle $\chi_P \in (-\pi, \pi]$ and pursuer elevation angles $\gamma_P \in [-\pi/2, \pi/2]$, while $F_T$ is obtained by rotation of $F_L$ through the target's azimuth angle $\chi_T \in (-\pi, \pi]$ and the target's elevation angles $\gamma_T \in [-\pi/2, \pi/2]$. 

Assuming the vehicles to be point masses with nonholonomic constraints, the equations of the relative motion between the pursuer and the target are governed by
\begin{subequations}
\label{eqn:dyna}
\begin{align}
\label{eqn:rel_dyn}
    \dot{r} &= V_T\cos\gamma_T\cos\chi_T-V_P\cos\gamma_P\cos\chi_P,   \\ 
    \dot{\theta} &= \frac{V_T\sin\gamma_T-V_P\sin\gamma_P}{r},\\
    \dot{\psi} &= \frac{ V_T\cos\gamma_T\sin\chi_T-V_P\cos\gamma_P\sin\chi_P}{r\cos\theta},
\end{align}
\end{subequations}
where $[V_i, \gamma_i$, $\chi_i]^\top$ denote the states of the respective vehicle. The rates of the vehicles' speed and turn are related as
\begin{subequations}
\label{eqn:rtspang}
    \begin{align}
        \dot{\gamma}_i &=\frac{a_i^\gamma}{V_i} -\dot{\psi}\sin\theta\sin\chi_i-\dot{\theta}\cos\chi_i, \label{eqn:kine_gamp}\\
    \dot{\chi}_i &= \frac{a_i^\chi}{V_P\cos\gamma_i}  + \dot{\psi}\tan\gamma_i\cos\chi_i\sin\theta-\dot{\psi}\cos\theta \nonumber \\
    &~~~-\dot{\theta}\tan\gamma_i\sin\chi_i,\label{eqn:chi_Pdot}\\
    \dot{V}_i &=a_i^r, \label{eqn:radialkin}
    \end{align}
\end{subequations}
where $a_i^r$ denotes the vehicle's radial acceleration, $a_i^{\gamma}$ denotes the lateral acceleration component in the pitch plane, and $a_i^{\chi}$ denotes the lateral acceleration component of the vehicles in the yaw plane such that the subscript $i \in \{P, T\}$. The accelerations $\mathbf{u}_T=[a_T^r, a_T^{\gamma}, a_T^{\chi} ]^\top$  and $\mathbf{u}_P=[a_P^r, a_P^{\gamma}, a_P^{\chi} ]^\top$ represent their respective control inputs in their respective body-fixed frames $F_P$ and $F_T$.  As evident from \eqref{eqn:radialkin}, the radial accelerations $a_i^r$ directly influence the vehicles' speeds, while the lateral acceleration components $a_i^{\gamma}$ and $a_i^{\chi}$ affect their turn rates, thereby influencing the direction of their velocities, as observed in \eqref{eqn:kine_gamp}-\eqref{eqn:radialkin}. Such a model provides a practical representation for aerial vehicles constrained to steer with a minimum turning radius and controlled through accelerations. The accelerations of the target are assumed to be bounded, that is, $\vert a_T^{r}\vert \leq a_{\max}^{r}$, $\vert a_T^{\gamma}\vert \leq a_{\max}^{\gamma}$ and $\vert a_T^{\chi}\vert \leq a_{\max}^{\chi}$, which is reasonable in practice. We articulate the assumption that the pursuer possesses a superior kinematic capability relative to the target, characterized by the inequalities $\max V_p(t)>\max V_T(t)$, $\max a_P^\gamma(t)>a_{\max}^\gamma$, and $\max a_P^\chi(t)>a_{\max}^\chi$. This implies that the pursuer's maximum allowable linear speed and lateral acceleration components exceed those of the target, thereby rendering more agility to the pursuer. It is also crucial to emphasize that the kinematics described in \eqref{eqn:dyna} is solely based on relative variables. Such an approach is particularly advantageous in GPS-denied environments, as it relies exclusively on relative measurements. Hence, we assume that the pursuer is equipped with onboard sensors to collect relative information. This setup eliminates the necessity for costly vehicle-to-vehicle communication and allows us to design an algorithm that operates effectively with limited information.
\subsection{Design of control objectives}
Our objective is to design the control inputs for the pursuer to enclose a mobile target while ensuring that the pursuer and the target do not collide ($r>0\; \forall\; t\geq0$). Meanwhile, we also aim to minimize the pursuer's overall energy expenditure. To this end, rather than tracking the target's speed, we simply let the pursuer attain a desired linear speed $V_d$, which is predetermined and remains constant throughout the engagement. Thus, $\lim_{t\to\infty}V_P\to V_d$ is the first objective. Additionally, a necessary step in target enclosing entails pursuer converging to a fixed desired proximity $r_d$ from the target, that is, $\lim_{t\to\infty}r\to r_d$. 

\begin{figure*}[ht!]
    \centering    \includegraphics[width=0.40\linewidth]{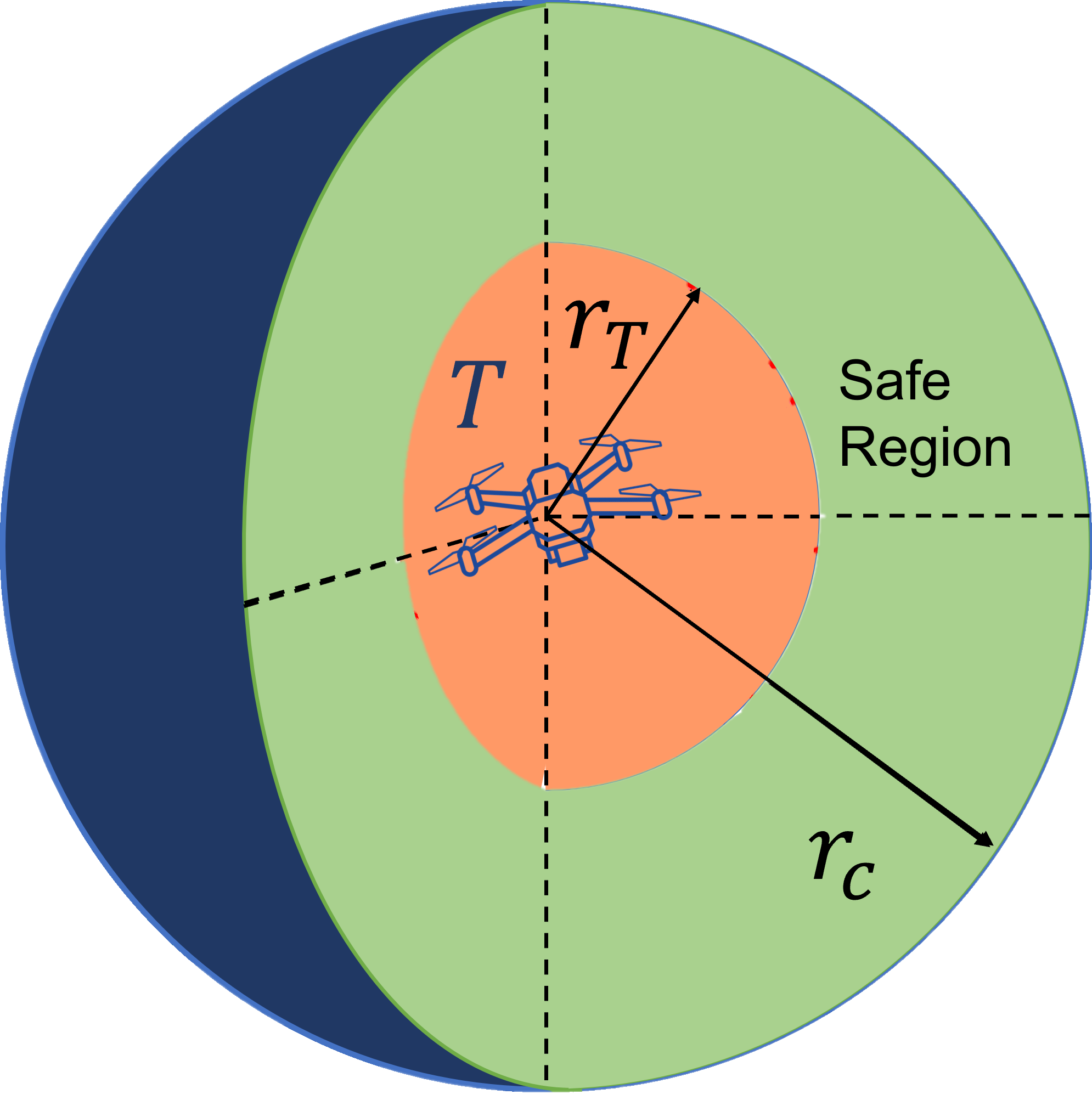}
    \caption{Pursuer's regions of operation.}
    \label{fig:reg}
\end{figure*}

 It is worth mentioning that most of the prior works on target enclosing, e.g., \cite{LAN2010381,doi:10.2514/1.G006403,doi:10.2514/1.G006957,9924233,doi:10.2514/1.G007539,ranjan2024selforganizing,doi:10.2514/6.2024-0124}, only satisfy the second objective to attain target enclosing wherein the pursuer converges to a predetermined orbit around the target, without considerations of explicit safety (that is, ensuring $r>0,\; \forall\;t\geq0$) in the design. In practice, an agile maneuvering target may prevent the pursuer from maintaining a position at the desired proximity, potentially resulting in a collision with the target. Therefore, our aim is to design a guidance law that guarantees pursuer-target collision avoidance at all times. The safety constraint utilized in our approach is outlined as $r\in[r_T,r_C]$,
where $r_{T}$ denotes the radius of the threat region, which may be determined based on the situation and actual physical parameters such as minimum turn radius of the pursuer, minimum collision avoidance distance, sensing radius, lethal radius, or range of target weapons. On the other hand, $r_{C}$ denotes the radius of the connectivity region that may be selected based on the sensors utilized for detecting and tracking the target. The inequality $r \geq r_{T}$ ensures that the pursuer and the target maintain a non-zero distance between them throughout the engagement, whereas $r \leq r_{C}$ represents an additional safety consideration to ensure that the pursuer does not maneuver very far away from the target rendering it incapable of enclosing the target.  Therefore, the pursuer satisfying such safety constraints confines its movement within the safe region of operation as depicted in \Cref{fig:reg}. 

Let us denote the configuration or the position of the pursuer with respect to the target by $\phi(t)=[r(t), \theta(t), \psi(t)]^\top$. The safe region of operation consists of a spherical shell (or volume of cloud surrounding the target) of a fixed thickness $r_C-r_{T}$, which represents the allowable positions of the pursuer that ensures safety while enclosing the target. This safe region is defined by the set, $\Omega_p\coloneqq \Big\{\left(r,\theta,\psi\right)\big\vert\, r\in\left[r_{T},r_{C}\right], \theta \in [-\pi/2, \pi/2], \psi \in[-\pi, \pi) \Big\}$. Therefore, flexible target enclosing is achieved if the pursuer's relative position remains within a predefined set $\Omega_P$, that is, $\phi(t) \in \Omega_P ,\; \forall t\geq0$. This set $\Omega_p$  remains constant in $F_L$ but varies with time in $F_I$ due to the mobility of the target. The above condition ensures that every relative position pertaining to $\Omega_P$ is a stable position for the pursuer to maintain enclosing since the pursuer is guaranteed to remain bounded within $\Omega_P$. Our approach offers the pursuer multiple options for 3D orbits, enhancing its ability to effectively enclose the target since the pursuer is now permitted to occupy a spherical shell representing the safe region. This is unlike maintaining a fixed proximity to the target, where the pursuer can take only circular/elliptical orbits. 
 
 In this letter, we aim to develop a generalized 3D guidance law for the pursuer to flexibly enclose an arbitrarily maneuvering target and guarantee the safety of the pursuer from collision with the target. The proposed guidance strategy enables the pursuer to maintain a fixed linear speed $V_d$ to lower the energy expenditure and converge to a set of relative positions $\Omega_p$. Further, we do not decouple the pursuer's motion into separate channels, allowing it to maneuver freely in 3D, converging to any orbit situated surrounding the target within a volume of cloud (around $r_d$). To summarize, the primary control objectives are to ensure (i) $\lim_{t\rightarrow\infty} V_P(t)\rightarrow V_d$, (ii) $\lim_{t\rightarrow\infty} r(t)\rightarrow r_d$, (iii) $\phi(t)\in\Omega_p\;\forall\;t\geq0$. Additionally, we aim to minimize the pursuer's overall control effort during enclosing by weighted effort allocation, thereby providing it even greater flexibility to shape its trajectory. Before discussing the main results, we present an important result on the second-order Barrier Lyapunov Function relevant to control design with constraints on state variables. Such a function ensures stability with output constraints to guarantee the boundedness of the related error variable.
\begin{lemma}\cite{TEE2009918}
\label{eqn:lemm_lyapunov}
    Consider the open sets, $\mathcal{Z}_1 \coloneqq \{z_1 \in \mathbb{R}, -a<z_1<b\} \subset \mathbb{R}$,  $\mathcal{N}\coloneqq \mathbb{R} \times \mathcal{Z}_1 \subset \mathbb{R}^{2}$ and the system $\dot{\eta} = h(t,\eta)$,
    where $\eta \coloneqq[w, z_1]^\top \in \mathcal{N}$ and $h$ is piecewise continuous in $t$ and locally Lipschitz in z, uniform in t, such that $h: \mathbb{R}_{>0} \times \mathcal{N} \rightarrow \mathbb{R}^{2}$. Suppose there exist two continuously differentiable and positive definite functions $U:\mathbb{R} \rightarrow \mathbb{R}_{\geq0}$ and $V_1$, such that
    $ V_1(z_1) \rightarrow \infty, \; \text{as}\;z_1 \rightarrow -a\; \text{or} \;z_1 \rightarrow b$ and $\gamma_1\left(\lVert w \rVert\right)\leq U(\lVert w \rVert) \leq \gamma_2 \left(\lVert w \rVert\right)$,
    where $\gamma_1$ and $\gamma_2$ are $K_\infty$ functions. Let $V_2(\eta) = V_1(z_1)+U(w)$ and $\epsilon(0)$ belong to the set $\epsilon \in (-a, b)$. If the given inequality holds, $\dot{V} =\frac{\partial V}{\partial\eta}h\leq0 \nonumber$,
    then $\epsilon(t)$ remains in the open set $\epsilon \in (-a, b),\; \forall \, t \in [0,\infty).$
\end{lemma}

\section{Main Results}
In this section, we design the radial and lateral accelerations of the pursuer to satisfy the control objectives. We demonstrate that the proposed radial acceleration is independent of the lateral acceleration to simplify the design and lower the energy expenditure. This, in turn, will also aid in lowering the lateral acceleration effort because it may exhibit some dependence on the radial acceleration. This will be shown later.
\subsection{Design of the pursuer's linear speed}
We first design the pursuer's radial acceleration to maintain the desired linear speed $V_d$, while enclosing the mobile target. To this end, we define the error as, $e_v = V_P - V_d$.
On differentiating $e_v$ with respect to time and using \eqref{eqn:radialkin}, we obtain the dynamics of this error variable as $\dot{e}_v =\dot{V}_P=a_P^r$,
since $\dot{V}_d=0$. From this equation, it is evident that the linear speed error rate is directly related to the radial acceleration. Therefore, we propose the pursuer's radial acceleration as
    \begin{align}
    \label{eqn:rad_ctrl}
        a_P^r = -K_v(V_P-V_d),
    \end{align}
where $K_v >0$ denotes the controller gain.
\begin{theorem}
    \label{eqn:ctrl_ar}
    Consider the equations of pursuer-target relative motion \eqref{eqn:dyna}. The pursuer's radial acceleration  \eqref{eqn:rad_ctrl} allows it to maintain a constant speed throughout the engagement.
\end{theorem}
\begin{proof}
    Consider a Lyapunov Function candidate  $V_v =\frac{1}{2}e_v^2$. On  differentiating $V_v$ with respect to time and substituting for $\dot{e}_v$, we obtain, $\dot{V}_v=e_v\dot{e}_v=e_va_P^r$.
    Choosing $a_P^r$ as in \eqref{eqn:rad_ctrl} renders the time derivative of the Lyapunov function candidate as $\dot{V}_v = -K_ve_v^2 <0 \, \forall \, e_v \in \mathbb{R} \setminus \{0\},$
    if $K_v>0$. This implies that under the proposed control law, $V_v$ and hence $e_v$ decays to zero, resulting in the pursuer's speed converging to the desired speed.  On substituting the proposed control law $a_P^r$, we obtain the closed loop error dynamics as $\dot{e_v} = -K_ve_v$,
    which upon integration yields $e_v(t) =e_v(0)e^{-K_vt}$, thereby showing asymptotic convergence ($e_v\to 0$ as $t \to \infty$).
\end{proof}
 It follows from \Cref{eqn:ctrl_ar} that $V_P \to V_d$ as $t \to \infty$. In comparison to previous works, our approach simplifies the linear speed control design to lower the energy expenditure by requiring the pursuer to maintain a constant speed as opposed to dynamically changing $V_P$.
Since we have assumed the pursuer to have a linear speed advantage over the target, selecting $V_d$ such that $V_d>\sup_{t\geq 0}V_T$, provides an appropriate choice of $V_d$ that may provide the pursuer the capability to enclose the target at all times. In practice, such information can be easily obtained using onboard sensors or a ground station.
\subsection{Design of the pursuer's range controller}
\label{sec:rangtctrl}
We now derive the lateral acceleration for the pursuer that allows it to converge to the desired proximity from the target while remaining in the safe region. We show that the proposed guidance law guarantees the global convergence of the range error with respect to the predefined state constraints. This implies that starting from all feasible initial configurations, the pursuer is able to achieve stable enclosing behavior without violating the safety constraints. Given the constant desired proximity $r_d$ of the pursuer from the target, we define the range error as $\epsilon(t)=r(t)-r_d$.
 Based on the safety constraint, the pursuer-target relative range $r$ is bounded between the predefined constraints in order to confine the movement of the pursuer within the safe region. Therefore, the error variable should also remain bounded as $-a < \epsilon < b$, 
where $a=r_d-r_{T}$ and $b= r_{C} - r_d$ denote the additional parameters representing the range error at boundaries. It is evident that different choices of $a$ and $b$ may lead to asymmetric bounds on the error variable. A larger value of $a$ will provide more flexibility to the pursuer to avoid collision with the target, while a smaller value of $b$ ensures that the pursuer does not maneuver far away from the desired proximity to ensure successful enclosing of the target at all times. 

To analyze the behavior of range error, we differentiate $\epsilon$ with respect to time and use \eqref{eqn:rel_dyn} to obtain the dynamics of the error variable as, $\dot{\epsilon} =\dot{r}-\dot{r}_d= V_T\cos\gamma_T\cos\chi_T-V_P\cos\gamma_P\cos\chi_P$
since $\dot{r}_d=0$. It can be observed that the range error rate is directly related to the relative velocity along the pursuer-target LOS. Additionally, we can infer that for a fixed speed $V_d$, the UAV can adjust $\gamma_P$ and $\chi_P$ to influence its velocity (direction) and thereby control its distance from the target. On further differentiating $\dot{e}_r$ with respect to time and using  \eqref{eqn:dyna} and \eqref{eqn:rtspang}, we obtain the dynamics of error rate as
\begin{align}
     \ddot{\epsilon} &= r\dot{\theta}^2 + r\cos\theta^2 \dot{\psi}^2 +a_P^r\cos\gamma_P\cos\chi_P+ \mathcal{U} +\Delta,  \label{eqn:eff_Sta2}
\end{align}
where we have separately clubbed the target's and the pursuer's control input-dependent terms. Here $\Delta =-a_T^r\cos\gamma_T\cos\chi_T-\sin\gamma_T\cos\chi_Ta_T^\gamma -\sin\chi_Ta_T^\chi$ denotes the target's control input dependent term
and $\mathcal{U}=\sin\gamma_P\cos\chi_Pa_P^\gamma+\sin\chi_Pa_P^\chi$ denotes the effective control input combining the pursuer's lateral acceleration components.
It can be observed from \eqref{eqn:eff_Sta2} that the range error dynamics has a relative degree of two with respect to the pursuer's lateral acceleration components. Further, we will circumvent the need for exact information of the target's control (that may be difficult to obtain in practice) by reasonably treating $\Delta$ as a bounded uncertainty. It is important to stress that the proposed guidance strategy is designed considering the kinematics of the relative motion between the pursuer and the target, as given in \eqref{eqn:rtspang}. For a different vehicle dynamics model we can obtain $\ddot{r} = r\dot{\theta}^2 + r\cos\theta^2 \dot{\psi}^2 + \mathcal{U}^{'} +a_P^r\cos\gamma_P\cos\chi_P+ \Delta^{'}$, where $\mathcal{U}^{'}$ denotes the new control input and $\Delta^{'}$ denotes the new uncertainty. This implies that when a different vehicle model is used, the fundamental structure of the range error dynamics \eqref{eqn:eff_Sta2} remains the same, and the proposed approach still remains valid, thereby introducing robustness in the design. Note that various choices of $a_P^\gamma$ and $a_P^\chi$ may lead to the same $\mathcal{U}$. Therefore, we first propose the effective control input $\mathcal{U}$ to nullify the range error and its rate, given by
\begin{align}
    \mathcal{U}=&-r\dot{\theta}^2 - r\cos^2\theta \dot{\psi}^2-K_v\left(V_p-V_d\right)\cos\gamma_P\cos\chi_P\nonumber\\&+\dot{\alpha}-{K}_2\sign{\left(\dot{r}-\alpha\right)} 
    -\left[\frac{q}{a^2-\epsilon^2}+\frac{\left(1-q\right)}{b^2-\epsilon^2}\right]\epsilon \label{eqn:ctrl},
\end{align}
where a sufficient condition on the controller gain $K_2$ is such that $K_2 > \vert a_{\max}^r\vert+\vert a_{\max}^\gamma \vert + \vert a_{\max}^\chi \vert$ and $q\left(\epsilon\right) =   
    \begin{cases}
    1,\;\text{if}\; \epsilon>0\\
    0,\;\text{if}\; \epsilon\leq0\\
    \end{cases}$ denotes a switching function with
$\alpha$ denoting the stabilizing function given as
\begin{align}
       \alpha &= -\bigr[q(b^2-\epsilon^2)+\left(1-q\right)(a^2-\epsilon^2)\bigr]{K}_1 \epsilon^3, \label{eqn:stab_fcn}  
\end{align}
where $K_1>0$ is also a controller gain. For brevity, we drop the argument of $q(\epsilon)$ from everywhere, and it will be understood that $q$ is a function of $\epsilon$.
\begin{theorem}
\label{thm:1}
    Consider the equations of pursuer-target relative motion \eqref{eqn:dyna}. The pursuer's effective control effort  \eqref{eqn:ctrl} allows it to converge to the desired proximity with respect to the target while invariably remaining within a predetermined safe region surrounding the target.
\end{theorem}
\begin{proof}
Consider an asymmetric Lyapunov Barrier Function candidate, $V_1(\epsilon)= q \log{\frac{b^2}{b^2-\epsilon^2}} +\left(1-q\right) \log{\frac{a^2}{a^2-\epsilon^2}}$,
 which is a convex combination of two barrier functions with only one of them activated at a particular instance of time, based on $q$. Let us assume a pseudo-variable $z=\dot{\epsilon}-\alpha$, where $\alpha$ is a stabilizing function to be designed. Differentiating $V_1$ with respect to time and using $\dot{\epsilon}=z+\alpha$, we obtain, $\dot{V}_1(\epsilon)= \Big(\frac{q}{b^2-\epsilon^2} + \frac{1-q}{a^2-\epsilon^2}\Big)\epsilon\left(z+\alpha\right).$ By selecting the the stabilizing function as in  \eqref{eqn:stab_fcn}, it follows that $\dot{V}_1=-K_1 \epsilon^4+
    \Big(\frac{q}{b^2-\epsilon^2} + \frac{1-q}{a^2-\epsilon^2}\Big)\epsilon z.$ Now let us consider a second Lyapunov function candidate, $V_2(z,\epsilon)= V_1 +\frac{1}{2}z^2,$ where the original Lyapunov candidate is augmented with $z$-dependent quadratic term. On differentiating $V_2$ with respect to time, we obtain     
    $\dot{V}_2 = -K_1 \epsilon^4+
    \Bigr[\frac{q\epsilon}{b^2-\epsilon^2}+\frac{\left(1-q\right)\epsilon}{a^2-\epsilon^2}+r\dot{\theta}^2+ r\cos\theta^2\dot{\psi}^2 -K_v(V_p-V_d)\cos\gamma_P\cos\chi_P + \mathcal{U} +\Delta-\dot{\alpha}\Bigr]z$
 after using $\dot{z}=\ddot{\epsilon}-\dot{\alpha}$ and substituting for $\dot{V}_1$, $\ddot{\epsilon}$  and $a_P^r$. Choosing the effective control input as specified in \eqref{eqn:ctrl} simplifies the preceding expression to,
  $\dot{V}_2 = -K_1 \epsilon^{4} + \left(-K_2\sign(z)+\Delta\right) z $, which consequently implies $\dot{V}_2 \leq -K_1 \epsilon^{4} -\left(K_2-\vert\Delta\vert \right)\vert z \vert$ and hence $\dot{V}_2\leq -\Big(K_2-\left(\vert a_{\max}^r\vert+\vert a_{\max}^\gamma\vert+\vert a_{\max}^\chi\vert\right)\Big)\vert z \vert-K_1 \epsilon^{4}$,
that reveals the sufficient condition provided after \eqref{eqn:ctrl} to ensure that $\dot{V}_2<0 \; \forall\; \{\epsilon,z\} \in \mathbb{R}^2 \setminus \{0,0\}$.
Therefore, $\epsilon \to 0$ and $z\to 0$ as $t\to \infty$. This results in the pursuer converging to the desired proximity around the target, that is, $r \to r_d$. It can be further deduced from \Cref{eqn:lemm_lyapunov}, that $a<\epsilon(t)<b, \; \forall\; t>0$, provided that $a<\epsilon(0)<b$. 
\end{proof}
From \eqref{eqn:stab_fcn}, one may observe that $\alpha \to 0$ as $\epsilon \to 0$. Therefore, from \Cref{thm:1}, we can also conclude that $\dot{r} \to 0$ as $t \to \infty$ since $z=\dot{r}-\alpha$. At steady-state, the pursuer converges to the desired proximity from the target, maintaining constant speed $V_d$, resulting in $\mathcal{U}\to-r_d\dot{\theta}^2 - r_d\cos\theta^2 \dot{\psi}^2$ from \eqref{eqn:ctrl} denoting the centripetal acceleration required to maintain position on a sphere of radius $r_d$. \Cref{thm:1} also provides guarantees on the boundedness of the range error, that is, $a<\epsilon<b$. This also implies $r_{T}<r(t)<r_{C}$ since $r_d$ is a constant, resulting in $\phi(t) \in\Omega_p \;\forall \; t \geq0$ given  $\phi(0) \in \Omega_p$. 

By design, the pursuer's guidance law is capable of handling uncertainties without tracking overshoot since the stringent requirement of making the range error zero is rather relaxed by letting it stay within certain bounds. However, if there are uncertainties that may lead to steady-state errors, then we have, $ \dot{z}=\ddot{\epsilon}-\dot{\alpha}=-K_2~\sign{(z)} + \Delta(t)=0$,  which implies that $z_{ss}= \frac{\vert\Delta\left(t\right)\vert\sign\left(\Delta\left(t\right)\right)}{K_2}$ and $ \vert z_{ss} \vert  <\dfrac{\vert\Delta(t)\vert}{K_2}$. This implies that selecting higher values of gain $K_2$ will reduce the steady-state error in $z$. Similarly, we can express the dynamics of the tracking error in the steady state as $ \dot{\epsilon}=z_{ss}+\alpha= \frac{\Delta\left(t\right)}{K_2}-\bigr[q(b^2-\epsilon^2)+\left(1-q\right)(a^2-\epsilon^2)\bigr]{K}_1 \epsilon^3=0$, providing the upper bound on the steady-state values of ${\epsilon}$ as, $\left(\frac{\vert\Delta\vert}{\max\left(b^2,a^2\right)K_1K_2}\right)^{\frac{1}{3}}$, which can be made arbitrary small by suitable choice of design parameters. 
 Therefore, the effects of $\Delta$ on the steady-state range error are relatively small since the proposed controller is designed using the notions of the Barrier Lyapunov function that ensures $-a<\vert\epsilon(t)\vert<b$. This means that the pursuer will always remain on a stable enclosing path even with a non-zero steady-state error, thereby exhibiting robustness to target-dependent uncertainties. In practice, the bounds on the steady-state value of the error are typically much smaller than the bounds of the error itself. Therefore, one may obtain the relationship, $ \left(\frac{\vert\Delta\vert}{\max\left(b,a\right)^2K_1K_2}\right)^{\frac{1}{3}}<\max\left(a,b\right),$
which allows for the design of the gains based on the parameters $a,b$. This demonstrates that the proposed enclosing strategy provides the pursuer with the flexibility to occupy any position within the safe region set $\Omega_p$, while letting it remain on stable enclosing trajectories around the target regardless of the bounded uncertainties.

Furthermore, although it might seem from \eqref{eqn:ctrl} that the proposed effective control may potentially grow unbounded if $\epsilon =a$ or $\epsilon =b$. However, such a situation never arises as confirmed in \Cref{thm:1} since the error variable always remains within the predefined constraints, never reaching the constraint boundaries. The first two terms in \eqref{eqn:ctrl} represent the centripetal acceleration required to maintain a circular orbit at the current $r$ around the target. The third term in \eqref{eqn:ctrl} represents the component of the radial acceleration along the LOS, indicating a relationship between the radial and lateral accelerations. The fifth and the sixth terms in \eqref{eqn:ctrl} aid in adjusting the pursuer trajectory toward the desired proximity. Finally, the last two terms in \eqref{eqn:ctrl} also ensure that the pursuer consistently stays within the safe region surrounding the target. 
\subsection{Optimal Allocation of Lateral Acceleration Components}
As previously mentioned, the effective control input $\mathcal{U}$ is a combination of the components of lateral acceleration in the pitch and yaw planes. This implies that infinite choices of $a_p^\gamma$ and $a_p^\chi$ can result in the same $\mathcal{U}$. Hence, we now formulate an optimization problem to distribute the control effort in the pitch and the yaw planes to achieve the effective lateral acceleration control $\mathcal{U}$ as proposed in \eqref{eqn:ctrl}. We define the cost function for optimization as $\mathcal{J} = \sqrt{\left(\frac{a_P^\gamma}{w_1}\right)^2 + \left(\frac{a_P^\chi}{w_2}\right)^2}$,
where $w_1,w_2 >0$ denote the weight parameters in the pitch and yaw channels, respectively. This cost function may represent the pursuer's energy expenditure or maneuverability in terms of its acceleration in the pitch and yaw plane when $w_1=w_2$.
\begin{theorem}\label{thm:tm3}
    The optimal values of the pursuer's lateral acceleration components in the pitch and yaw plane,
     $ a_P^\gamma = \frac{\sin\gamma_P\cos\chi_Pw_1^2\mathcal{U}}{(\sin\gamma_P\cos\chi_P)^2w_1^2+(\sin\chi_P)^2w_2^2}$ and $a_P^\chi = \frac{\sin\chi_Pw_2^2\mathcal{U}}{(\sin\gamma_P\cos\chi_P)^2w_1^2+(\sin\chi_P)^2w_2^2}$, minimizes the cost function $\mathcal{J}$ under the constraints \eqref{eqn:ctrl}.
   
\end{theorem}
\begin{proof}
Using the relationship $\mathcal{U}=\sin\gamma_P\cos\chi_Pa_P^\gamma+\sin\chi_Pa_P^\chi$ we can rewrite the cost function as, 
 \begin{align}
     \mathcal{J} = \sqrt{\left(\frac{a_P^\gamma}{w_1}\right)^2 + \left(\frac{\mathcal{U}-\sin\gamma_P\cos\chi_Pa_P^\gamma}{\sin\chi_Pw_2}\right)^2}. 
     \label{eqn:eff_cst}
 \end{align}
Taking the partial derivative of the above equation with respect to $a_P^{\gamma}$, we get,
\begin{align}
     \frac{\partial \mathcal{J}}{\partial a_P^\gamma }= \frac{1}{\mathcal{J}}\Biggl(\frac{a_P^\gamma}{w_1^2}&-\left(\frac{\sin\gamma_P\cos\chi_Pa_P^\gamma}{\sin\chi_Pw_2}\right)\nonumber\\&\left(\frac{U-\sin\gamma_P\cos\chi_Pa_P^\gamma}{\sin\chi_Pw_2}\right)\Biggl).\nonumber\\
 \end{align}
 Equating the above equation to zero yields the optimal value of the lateral acceleration component in the yaw plane as,
 \begin{align}    a_P^{\gamma}=\left(\frac{\sin\gamma_P\cos\chi_Pw_1^2}{(\sin\gamma_P\cos\chi_P)^2w_1^2+(\sin\chi_P)^2w_2^2}\right)\mathcal{U}. \label{eqn:gam_ctrl}
 \end{align}
 Evaluating the second partial derivative of the cost function with respect to $a_P^\gamma$, we obtain,
 \begin{align}
     \frac{\partial^2 \mathcal{J}}{\partial^2 a_P^\gamma } =\frac{1}{\mathcal{J}}\left(\frac{1}{w_1^2}+\frac{\left(\sin\gamma_P\cos\chi_P\right)^2}{\left(\sin\chi_Pw_2\right)^2}\right).
 \end{align}
 It is evident from the above equation that the second derivative is always greater than zero, which implies that the control input \eqref{eqn:gam_ctrl}, minimizes the cost $\mathcal{J}$. Similarly rewriting the cost function in terms of $a_P^\chi$ and following the above procedure, we obtain the optimal value of the lateral acceleration component in the yaw plane as,
 \begin{align}
 a_P^{\chi}=\left(\frac{\sin\chi_Pw_2^2}{(\sin\gamma_P\cos\chi_P)^2w_1^2+(\sin\chi_P)^2w_2^2}\right)\mathcal{U} \label{eqn:chi_ctrl}
 \end{align}
 This concludes the proof.
\end{proof}
It may seem that the acceleration components in \Cref{thm:tm3} may grow unbounded if $(\sin\gamma_P\cos\chi_P)^2w_1^2+(\sin\chi_P)^2w_2^2 = 0$. However, this expression can become zero if and only if $\gamma_p=0$ and $\chi_p=0$. We show that such a scenario rarely arises and does not affect the target enclosing behavior. To analyze the dynamics of the pursuer's angles, we define the pursuer's effective lead angle $\sigma_P$ as the angle that the velocity vector subtends to the LOS and is given by the relationship, $\cos\sigma_P =\cos\gamma_P\cos\chi_P$.
Thereby, we can infer that  $\sigma_P\to 0$ if both $\gamma_P\to 0$ and $\chi_P\to0$. Therefore, showing that $\sigma_P=0$ is not an equilibrium point is sufficient to show that the components in \Cref{thm:tm3} do not grow unbounded. To this end, we differentiate the above-mentioned lead angle relationship with respect to time and use $a_p^\gamma$ and $a_p^\chi$ from \Cref{thm:tm3} to obtain the dynamics of pursuer's lead angle at the steady state when $\gamma_p=0$, $\chi_P=0$ since $ \dot{\sigma}_P =\frac{-r_d\dot{\theta}^2 - r_d\cos\theta^2 \dot{\psi}^2}{\sin\sigma_P},$
where we have substituted $\epsilon\to0$, $\dot{\epsilon}\to0$, $\alpha \to 0$ and $\dot{\alpha}\to0$. From this, it is evident that when $\sigma_P\to0$, $\dot{\sigma}_P \to \infty$, which shows that $\sigma_P=0$ is not an equilibrium point. However, in the transient phase, $\sigma_P \to 0$ may occur momentarily and result in the saturation of lateral acceleration components (due to finite control inputs in practice) that steers the pursuer away from $\sigma_P=0$ immediately. Hence, the proposed lateral acceleration components in \Cref{thm:tm3} are non-singular almost everywhere. 
Furthermore, the proposed radial acceleration \eqref{eqn:rad_ctrl} and the optimal lateral acceleration components (as in \Cref{thm:tm3}) for the pursuer are dependent only on relative variables and independent of target control inputs, making our design robust to target's maneuver and lucrative for the environment where global information is not readily available. 

\section{Simulation Results}
\begin{figure*}[ht!]
    \centering    \includegraphics[width=0.8\linewidth]{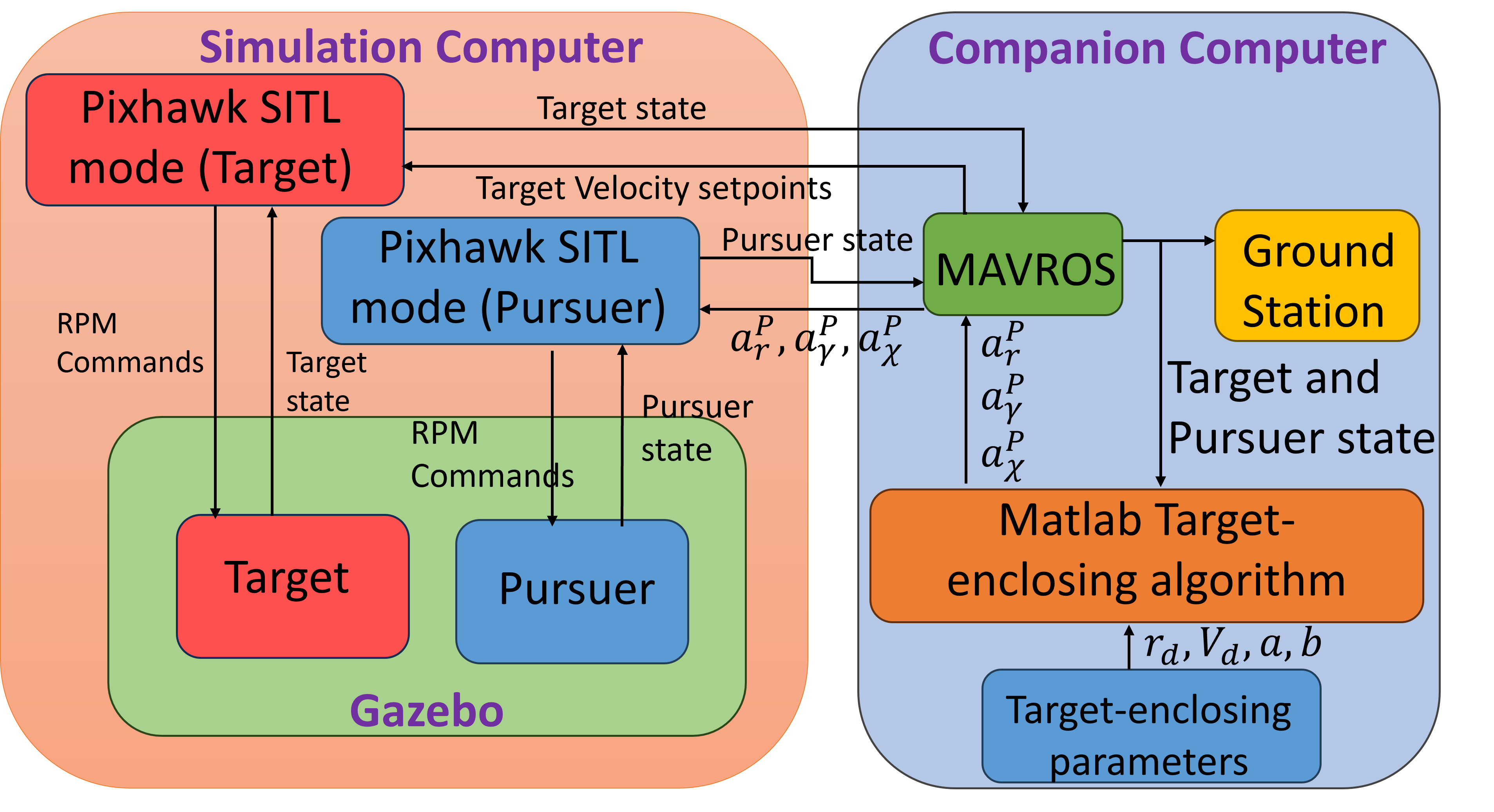}
    \caption{SITL Setup.}
    \label{fig:sitl}
\end{figure*}

To demonstrate the efficacy of the proposed flexible target enclosing guidance, which provides the pursuer robustness to uncertainties and safety from the collision with the target, we implement the guidance laws for various vehicle dynamical models and against different target maneuvers. In the results that follow, P-Kinematic pertains to simulation results when the guidance laws are implemented on the pure kinematic model \eqref{eqn:rtspang}. P-Uncertain implies that the guidance laws are implemented on the pure kinematics with an added disturbance of $1\sin(t)$ in the plant dynamics \eqref{eqn:rtspang}, whereas P-SITL represents the results when the guidance laws are implemented on a high-fidelity 6DOF quadrotor model in software-in-the-loop (SITL) simulation. The SITL simulation replicates conditions closer to real-world scenarios, emulating a high-fidelity quadrotor model and incorporating various components (autopilot, ground station, companion computer) essential for autonomous flight. Our SITL setup employs Gazebo software to simulate a high-fidelity quadrotor model, Pixhawk autopilot, and MAVROS node of the Robot Operating System (ROS), connecting the vehicles to a companion computer \cite{Fvr}. Further, MATLAB is employed in the companion computer to receive vehicle information and issue the proposed guidance commands. Given our primary objective of developing the guidance law, we allow the pursuer access to the target's global information. This facilitates the computation of requisite relative information for calculating the control inputs. Our simulation setup, operating on a 16 GB RAM AMD Ryzen 7 PC, achieves a sampling rate of 0.05 seconds (20Hz) for the guidance loop. \Cref{fig:sitl} depicts the flow of information between different components of the SITL setup.  

Further, to demonstrate robustness against target maneuver, we consider three distinct target scenarios. In the first scenario, the target remains stationary (ST), while in the second scenario, the target moves in a straight line at a slow constant speed of $V_T=2$ m/s (CVT).  Furthermore, in the third scenario, the target performs high-speed arbitrary maneuvers (MT), following a velocity profile described by $\mathbf{v_T} (t)= \left[3.5,1.5\sin\left(\dfrac{4\pi}{100}t\right), \sin\left(\dfrac{8\pi}{100}t\right)\right]^\top$ in the inertial frame of reference. Therefore, in the third scenario, the target's speed is time-varying such that $1<v_T<6$ . For the simulations involving slower target speeds (that are the first and the second scenario, respectively), the pursuer's desired proximity and speed are selected as $r_d =8$ m and $V_d=5$ m/sec, respectively.  For the third scenario (higher target speed), the pursuer's desired proximity and speed are chosen as $r_d =12$ m and $V_d=8$ m/sec.
For all the experiments, the pursuer starts at $[x_P, y_P, z_P]^\top$ $=[0, 0, 15]^\top$ m, with $\gamma_P = 10^\circ$ and $\chi_P = 10^\circ$, while the target starts at $[x_T, y_T, z_T]^\top$ $= [12, 12, 15]^\top m$, with $\gamma_T = 10^\circ$ and $\chi_T = 10^\circ$, such that the initial values of relative variables are given as $r = 18 $ m, $\theta = 0^\circ$ and $\psi = 45^\circ$. The controller parameters and gains are chosen as $a = 5$ m, $b=15$ m, $K_1=0.008$, $K_2=30$, $w_1=0.5$ and $w_2=0.5$. In the trajectory plots that follow, circular markers represent the initial position of the vehicles. Additionally, the videos of the vehicles' trajectories for the simulations presented in this paper can be found on \href{https://youtu.be/UU704o_966s}{https://youtu.be/UU704o\_966s}. 

\begin{figure*}[ht!]
     \centering
     \begin{subfigure}[t]{0.49\linewidth}
         \includegraphics[width=\linewidth]{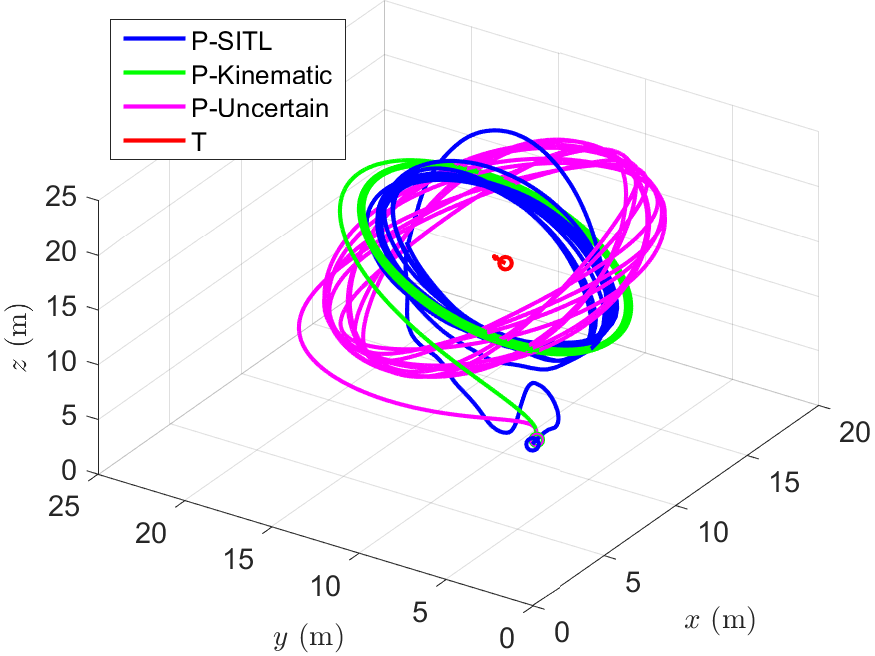}
         \caption{ST.}
         \label{fig:traj_sta}
         \end{subfigure}
         \begin{subfigure}[t]{0.49\linewidth}
         \includegraphics[width=\linewidth]{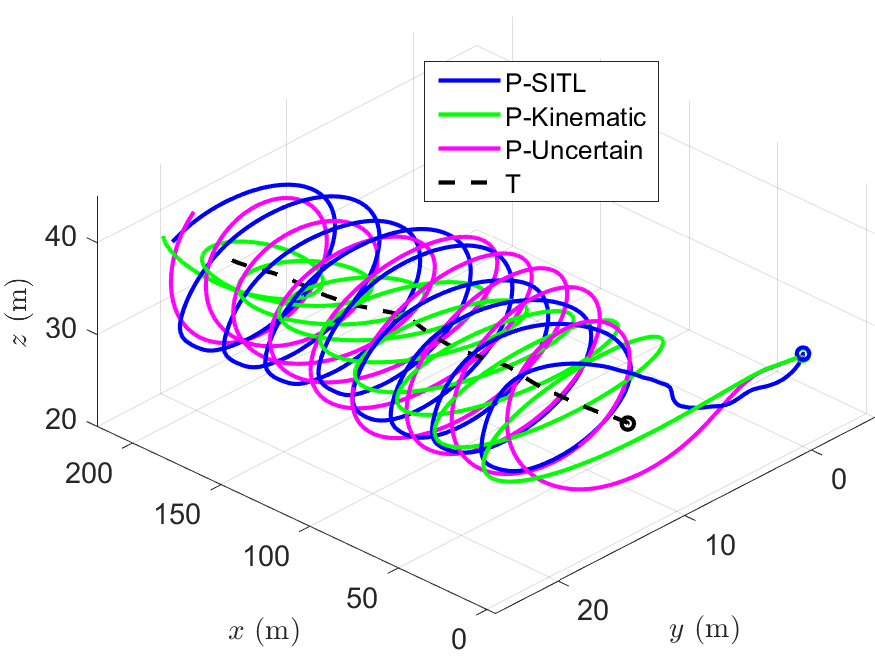}
         \caption{CVT.}
         \label{fig:traj_stline}
         \end{subfigure}
         \begin{subfigure}[t]{0.49\linewidth}
         \includegraphics[width=\linewidth]{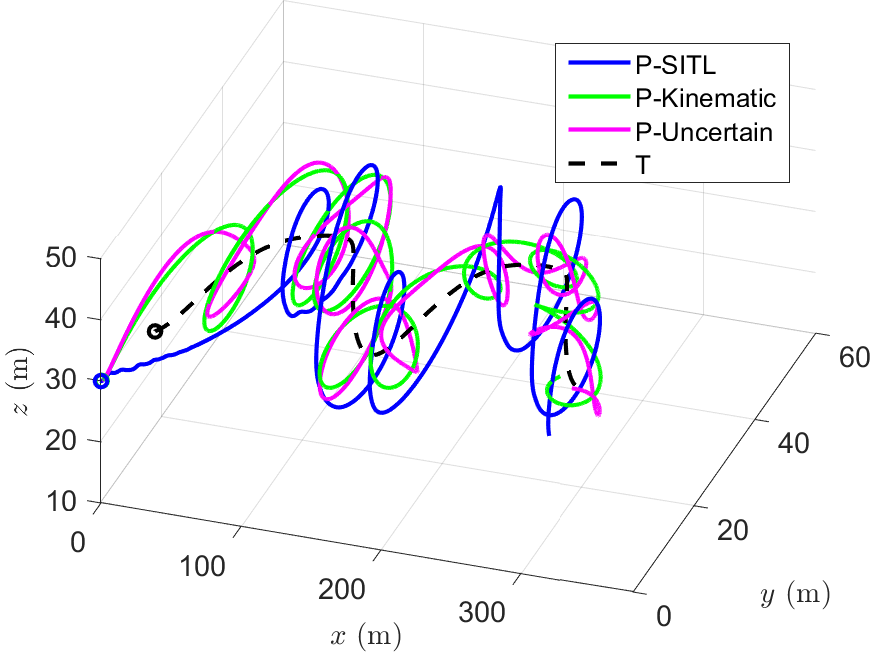}
         \caption{MT.}
         \label{fig:traj_arb}
         \end{subfigure}     
     \caption{Pursuer's trajectories for various target maneuvers.}
     \label{fig:traj}
 \end{figure*}

\begin{figure*}[h]
     \centering
     \begin{subfigure}[t]{0.32\linewidth}
         \includegraphics[width=\linewidth]{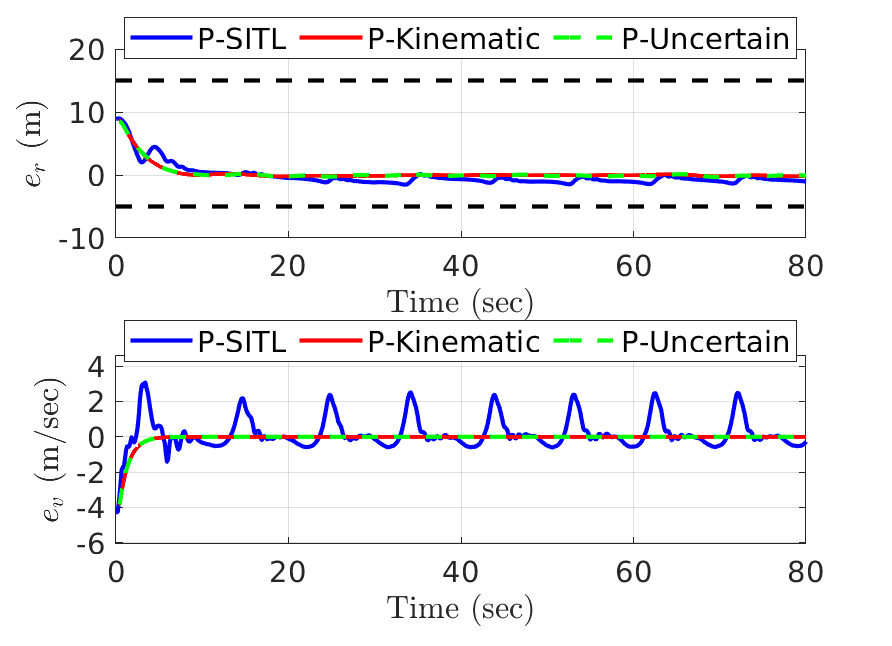}
         \caption{ST.}
         \label{fig:err_sta}
         \end{subfigure}
         \begin{subfigure}[t]{0.32\linewidth}
         \includegraphics[width=\linewidth]{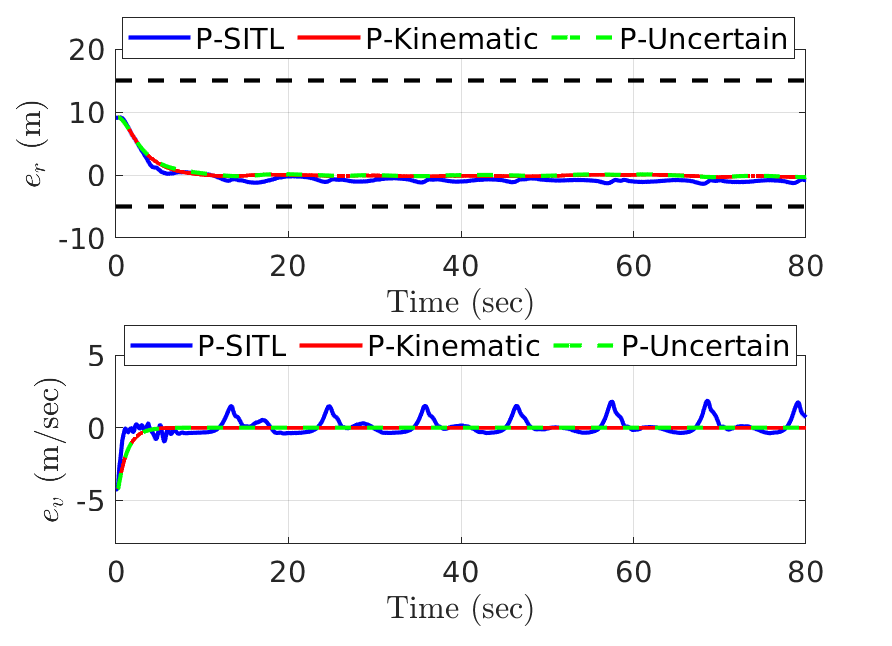}
         \caption{CVT.}
         \label{fig:err_stline}
         \end{subfigure}
         \begin{subfigure}[t]{0.32\linewidth}
         \includegraphics[width=\linewidth]{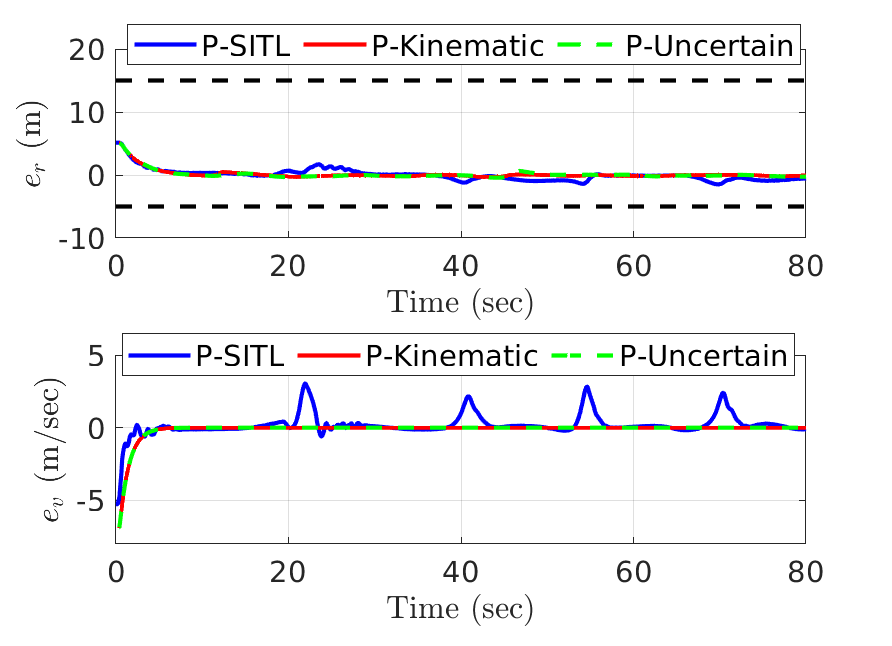}
         \caption{MT.}
         \label{fig:err_arb}
         \end{subfigure}     
     \caption{Pursuer's error variables for various target maneuvers.}
     \label{fig:terr}
 \end{figure*}

\begin{figure*}[h]
     \centering
     \begin{subfigure}[t]{0.32\linewidth}
         \includegraphics[width=\linewidth]{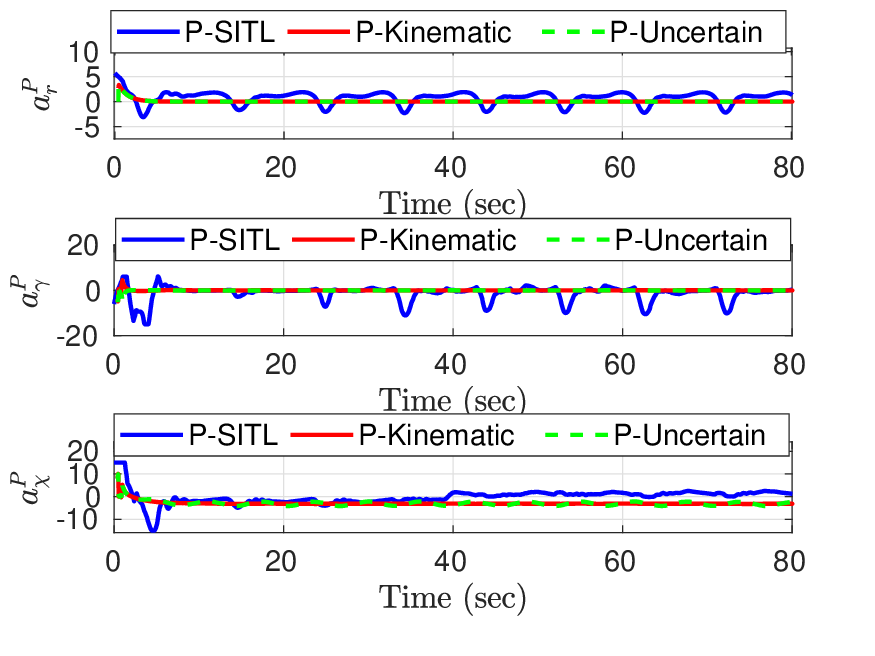}
         \caption{ST.}
         \label{fig:ctrl_sta}
         \end{subfigure}
         \begin{subfigure}[t]{0.32\linewidth}
         \includegraphics[width=\linewidth]{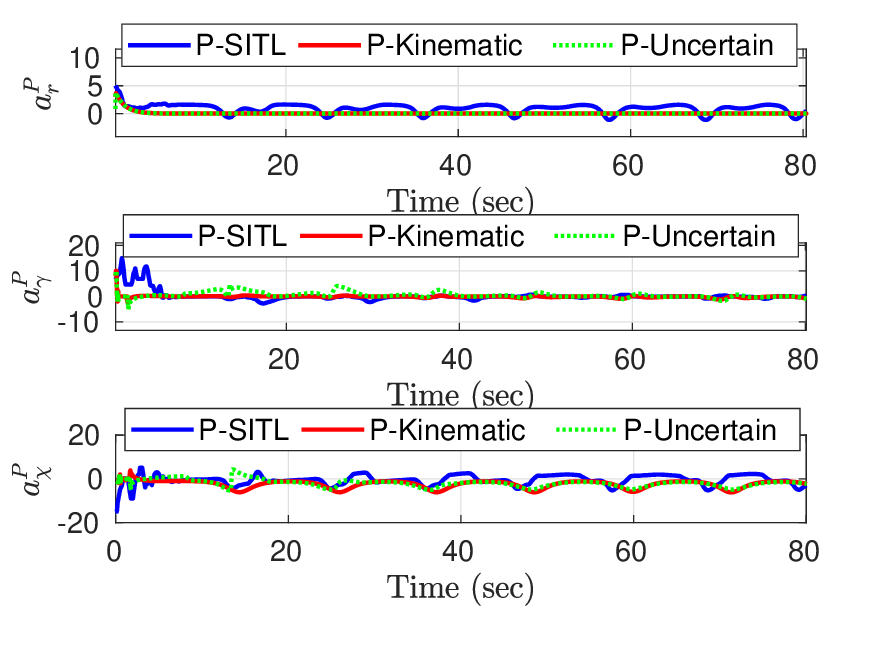}
         \caption{CVT.}
         \label{fig:ctrl_stline}
         \end{subfigure}
         \begin{subfigure}[t]{0.32\linewidth}
         \includegraphics[width=\linewidth]{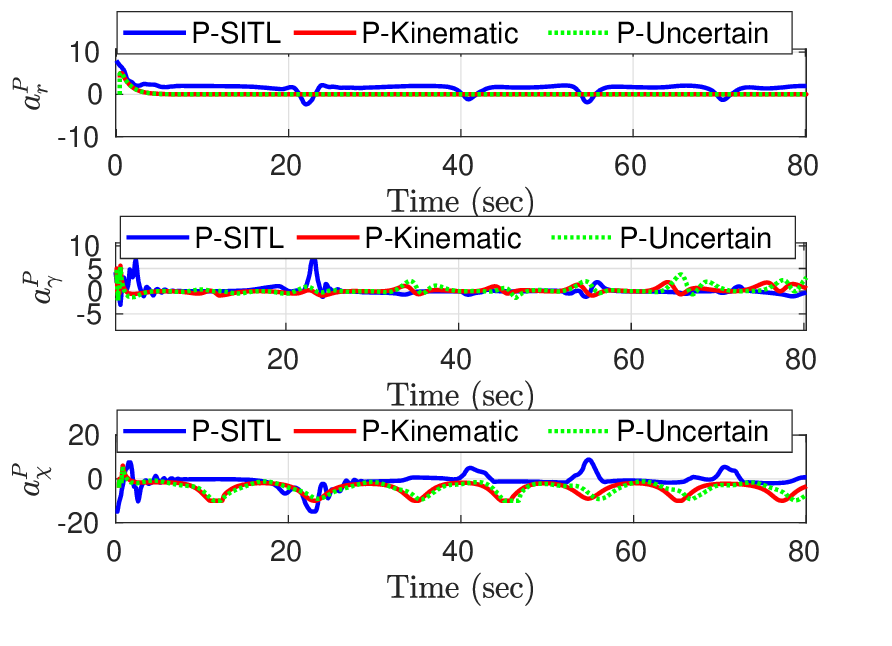}
         \caption{MT.}
         \label{fig:ctrl_arb}
         \end{subfigure}     
     \caption{Pursuer's control inputs for various target maneuvers.}
     \label{fig:ctrl_srb}
 \end{figure*}

\Cref{fig:traj} illustrates the pursuer's trajectories as it flexibly encloses the target, transitioning between different stable 3D orbits in various scenarios. For the stationary target, the flexibility of our guidance approach is most evident. In the P-Kinematic case, the pursuer quickly converges to a single 3D orbit. However, in the P-Uncertain case, the pursuer converges to the desired proximity but takes different enclosing orbits. In the P-SITL case, the pursuer follows 3D orbits with slight variations from the desired proximity. These variations occur because the dynamics in P-SITL is complex and close to real-world implementation, whereas P-Uncertain employs simplified approximate dynamics used to design the guidance laws. As observed, stable enclosing is still maintained in the presence of unmodeled dynamics, thereby providing robustness in the design. The error profiles of the pursuer's variables are shown in \Cref{fig:terr}, illustrating that the errors remain within the desired bounds, ensuring the pursuer stays in the safe region and successfully encloses the target in all scenarios. In both P-Uncertain and P-Kinematic cases, the range and speed errors quickly converge to zero, as the steady-state range error $\epsilon_{ss}\to0$ due to the small magnitude of any present uncertainty. However, for P-SITL, the error profiles show periodic deviations from zero, which occur when the UAV transitions between the highest and lowest points of the enclosing trajectory. These deviations are caused by unmodeled dynamics, leading to a higher level of uncertainty resulting in the pursuer switching between enclosing trajectories, as described in \Cref{sec:rangtctrl}. The profiles of the pursuer's control inputs are shown in \Cref{fig:ctrl_srb}, where for the P-Kinematics, P-Uncertain, and P-SITL, the control profiles converge to the required steady-state values for target enclosing, depicting robustness of the guidance law to the varying degree of variations of operating condition resulting in uncertainties. Particularly, the pursuer being able to enclose the target in P-SITL while also respecting the safety constraints demonstrates the guidance law exhibiting a certain degree of robustness during implementation.


\section{Conclusions}
We proposed a 3D guidance law for the pursuer to enclose a mobile target where the pursuer is constrained to maneuver only within the safe region of operation. The pursuer's radial acceleration is designed to achieve the desired speed, while the lateral accelerations are designed to steer the pursuer on enclosing shapes around the target. Our design provides guarantees on the boundedness of the pursuer-target relative range between the predefined bounds, preventing pursuer-target collision. We utilized coupled engagement kinematics and allocated control effort via static energy minimization, resulting in generalized 3D safe target enclosing behavior for the pursuer, which only requires information on relative variables. Although our design does not account for aerodynamic parameter variation or external/measurement disturbances, implementation of the guidance law on a UAV in SITL simulations depicted the robustness of our approach to target maneuver, vehicle dynamics, and autopilot dynamics in achieving stable enclosing behavior. Extension to multiagent settings, design with full dynamics-based vehicle modeling, and including additional constraints such as obstacles could be interesting future directions to pursue. 

\bibliographystyle{ieeetr}
\bibliography{references}

 \end{document}